\documentclass{article}[12pt]
\usepackage{epsfig,subfigure}
\newcommand{\be}{\begin{equation}}
\newcommand{\ee}{\end{equation}}
\newcommand{\bea}{\begin{eqnarray}}
\newcommand{\eea}{\end{eqnarray}}
\newcommand{\bc}{\begin{center}}
\newcommand{\ec}{\end{center}}
\newcommand{\bt}{\begin{tabular}}
\newcommand{\et}{\end{tabular}}
\newcommand{\bfig}{\begin{figure}}
\newcommand{\efig}{\end{figure}}
\newcommand{\bi}{\begin{itemize}}
\newcommand{\ei}{\end{itemize}}
\newcommand{\bleft}{\begin{flushleft}}
\newcommand{\eleft}{\end{flushleft}}
\newcommand{\bright}{\begin{flushright}}
\newcommand{\eright}{\end{flushright}}
\newcommand{\bpage}{\begin{minipage}}
\newcommand{\epage}{\end{minipage}}
\newcommand{\noi}{\noindent}

\newcommand{\pip}{\ensuremath{\pi^+\,}}
\newcommand{\pim}{\ensuremath{\pi^-\,}}
\newcommand{\piz}{\ensuremath{\pi^0\,}}
\newcommand{\Ks}{\ensuremath{K_S \,}}
\newcommand{\Kl}{\ensuremath{K_L\,}}

\newcommand{\Eta}{\ensuremath{\eta\,}}
\newcommand{\etap}{\ensuremath{\eta'\,}}
\newcommand{\phot}{\ensuremath{\gamma\,}}
\newcommand{\tp}{\ensuremath{\vartheta_P \,}}

\newcommand{\fip}{\ensuremath{\varphi_P\,}}
\newcommand{\dafne}{ DA\ensuremath{\Phi}NE }
\renewcommand{\to}{\ensuremath{\rightarrow}}
\newcommand{\cm}{\ensuremath{\,{\rm cm}}}

\newcommand{\s}{\ensuremath{\,{\rm s}}}
\newcommand{\ps}{\ensuremath{\,{\rm ps}}}
\newcommand{\pb}{\ensuremath{\,{\rm pb}}}
\newcommand{\pbinv}{\ensuremath{\,{\rm pb}^{-1}}}
\newcommand{\GeV}{\ensuremath{\,{\rm GeV}}}
\newcommand{\MeV}{\ensuremath{\,{\rm MeV}}}
\newcommand{\fikskl}{\ensuremath{\phi\rightarrow\Ks\Kl}}
\newcommand{\fipippimpiz}{\ensuremath{\phi\rightarrow\pi^+\pi^-\pi^0}}
\newcommand{\fietag}{\ensuremath{\phi\rightarrow\eta\gamma\;}}
\newcommand{\fietapg}{\ensuremath{\phi\rightarrow\eta'\gamma\;}}
\newcommand{\etapippimpiz}{\ensuremath{\eta\rightarrow\pip\pim\piz}}

\newcommand{\etagg}{\ensuremath{\eta\rightarrow\gamma\gamma\;}}
\newcommand{\etappippimeta}{\ensuremath{\etap\rightarrow\pip\pim\eta}}

\title{
\begin{flushright}
\small{
Contributed paper to Lepton Photon 2001 \\
Rome, July 23-28.}
\end{flushright}
Detection of $\fietapg$ ,
  $\fietag$ in $\pip\pim\gamma\gamma\gamma$ final state with KLOE at DA$\Phi$NE}
\date{ }
\author{The KLOE Collaboration}
\begin{document}
\maketitle

%\input defs
%%%%%%%%%%%%%%%%%%%%%%%%%%%%%%%%%%%%%%%%%%%%%%%%%%%%%%%%
%\hsize=160mm\vsize=240mm
%\hoffset=0mm\voffset=0mm
\def\ifm#1{\relax\ifmmode#1\else$#1$\fi}
\def\eps{\ifm{\epsilon}} \def\epm{\ifm{e^+e^-}}
\def\rep{\ifm{\Re(\eps'/\eps)}}  \def\imp{\ifm{\Im(\eps'/\eps)}}  
\def\DAF{DA$\Phi$NE}  \def\sig{\ifm{\sigma}}
\def\gam{\ifm{\gamma}} \def\to{\ifm{\rightarrow}}
\def\pip{\ifm{\pi^+}} \def\pim{\ifm{\pi^-}}
\def\po{\ifm{\pi^0}} 
\def\pic{\ifm{\pi^+\pi^-}} \def\pio{\ifm{\pi^0\pi^0}} 
\def\ks{\ifm{K_S}} \def\kl{\ifm{K_L}} \def\kls{\ifm{K_{L,\,S}}} 
\def\ksl{\ifm{K_S,\ K_L}} \def\ko{\ifm{K^0}}
\def\K{\ifm{K}} \def\LK{\ifm{L_K}}
\def\Kb{\ifm{\rlap{\kern.3em\raise1.9ex\hbox to.6em{\hrulefill}} K}}
\def\ab{\ifm{\sim}}  \def\x{\ifm{\times}}
\def\ff{$\phi$--factory}
\def\sta#1{\ifm{|\,#1\,\rangle}} 
\def\amp#1,#2,{\ifm{\langle#1|#2\rangle}}
\def\kob{\ifm{\Kb\vphantom{K}^0}}
\def\f{\ifm{\phi}}   \def\pb{{\bf p}}
\def\L{\ifm{{\cal L}}}  \def\R{\ifm{{\cal R}}}
\def\up#1{$^{#1}$}  \def\dn#1{$_{#1}$}
\def\etal{{\it et al.}}
\def\BR{{\rm BR}}
\def\radl{\ifm{X_0}}
\def\deg{\ifm{^\circ}} 
\def\th{\ifm{\theta}}
\def\To{\ifm{\Rightarrow}}
\def\ot{\ifm{\leftarrow}}
\def\fo{\ifm{f_0}} \def\epe{\ifm{\eps'/\eps}}
\def\pbrn{ {\rm pb}}  \def\cm{ {\rm cm}}
\def\mub{\ifm{\mu{\rm b}}} \def\s{ {\rm s}}
\def\RR{\ifm{{\cal R}^\pm/{\cal R}^0}}
\def\dt{ \ifm{{\rm d}t} } \def\dy{ {\rm d}y } \def\pbrn{ {\rm pb}}
\def\kp{\ifm{K^+}} \def\km{\ifm{K^-}}
\def\kkb{\ifm{\ko\kob}} 
\def\epe{\ifm{\eps'/\eps}}
\def\ppc{\ifm{\pi^+\pi^-}}
\def\ppo{\ifm{\pi^0\pi^0}}
\def\pppco{\ifm{\pi^+\pi^-\pi^0}}
\def\pppo{\ifm{\pi^0\pi^0\pi^0}}
\def\vare{\ifm{\varepsilon}}
\def\etap{\ifm{\eta'}}

\def\pt#1,#2,{#1\x10\up{#2}}

%%%%%%%%%%%%%%%%%%%%%%%%%%%%%%%%%%%%%%%%%%%%%%%%%%%%%%%%
                                
\def\B{Bari}
\def\b{\rlap{\kern.2ex\up a}}
\def\O{IHEP}
\def\o{\rlap{\kern.2ex\up b}}
\def\Fr{Frascati}
\def\fr{\rlap{\kern.2ex\up c}}
\def\Ka{Karlsruhe}
\def\ka{\rlap{\kern.2ex\up d}}
\def\Le{Lecce}
\def\le{\rlap{\kern.2ex\up e}}
\def\Mo{Moscow}
\def\mo{\rlap{\kern.2ex\up f}}
\def\N{Napoli}
\def\n{\rlap{\kern.2ex\up g}}
\def\BE{Beer-Sheva}
\def\be{\rlap{\kern.2ex\up h}}
\def\co{\rlap{\kern.2ex\up i}}
\def\Pi{Pisa}
\def\pI{\rlap{\kern.2ex\up j}}
\def\Ra{Roma I}
\def\ra{\rlap{\kern.2ex\up k}}
\def\en{\rlap{\kern.2ex\up l}}
\def\Rb{Roma II}
\def\rb{\rlap{\kern.2ex\up m}$\,$}
\def\Rc{Roma III}
\def\rc{\rlap{\kern.2ex\up n}}
\def\su{\rlap{\kern.2ex\up o}}
\def\T{Trieste/Udine}
\def\t{\rlap{\kern.2ex\up q}}
\def\V{Virginia}
\def\v{\rlap{\kern.2ex\up r}}
\def\Z{Associate member}
\def\hsa{ \ }
\normalsize
\noindent
A.~Aloisio\n,\hsa
F.~Ambrosino\n,\hsa
%{\it A.~Andryakov\mo,\hsa}
A.~Antonelli\fr,\hsa 
M.~Antonelli\fr,\hsa 
%{\it F.~Anulli\fr,\hsa}
C.~Bacci\rc,\hsa
G.~Barbiellini\t,\hsa 
F.~Bellini\rc,\hsa
G.~Bencivenni\fr,\hsa 
S.~Bertolucci\fr,\hsa 
C.~Bini\ra,\hsa
% ...
%%%%

\vglue 2mm

\def\aff#1{Dipartimento di Fisica dell'Universit\`a e Sezione INFN, #1, Italy.}

{\baselineskip=12pt
\parskip=0pt
\parindent=0pt
\def\hsb{\hskip 2.8mm}

\leftline{\b\hsb \aff{\B}}
\leftline{\o\hsb Permanent address: Institute of High Energy Physics of Academica Sinica, 
Beijing, China.}
\leftline{\fr\hsb  Laboratori Nazionali di Frascati dell'INFN, Frascati, Italy.}
\leftline{\ka\hsb  Institut f\"ur Experimentelle Kernphysik, Universit\"at \Ka,
Germany.}
\leftline{\le\hsb \aff{\Le}}
\leftline{\mo\hsb Permanent address: Institute for Theoretical and Experimental Physics, Moscow,
Russia.}
\leftline{\n\hsb Dipartimento di Scienze Fisiche dell'Universit\`a e 
Sezione INFN, \N, Italy.}
%\leftline{\be\hsb Physics Department, Ben-Gurion University of the Negev,
%Israel.}
\leftline{\co\hsb Physics Department, Columbia University, New York, USA.}
\leftline{\pI\hsb \aff{\Pi}}
\leftline{\ra\hsb Dipartimento di Fisica dell'Universit\`a ``La Sapienza'' e Sezione INFN,
Roma, Italy}
\leftline{\en\hsb Dipartimento di Energetica dell'Universit\`a ``La Sapienza'', Roma, Italy.}
\leftline{\rb\hsb Dipartimento di Fisica dell'Universit\`a ``Tor Vergata'' e Sezione INFN,
Roma, Italy}
\leftline{\rc\hsb Dipartimento di Fisica dell'Universit\`a ``Roma Tre'' e Sezione INFN,
Roma, Italy}
%\leftline{\s\hsb Istituto Superiore di Sanit\`a and Sezione INFN, ISS, Roma,
%Italy.}
\leftline{\su\hsb Physics Department, State University of New York 
at Stony Brook, USA.}
%\leftline{\te\hsb School of Physics and Astronomy, Tel Aviv, Israel.}
\leftline{\t\hsb \aff{\T}}
\leftline{\v\hsb Physics Department, University of Virginia, USA.}
%\leftline{\rlap{\kern.2ex\up*}\hsb\Z}     
   }

\newpage
\begin{abstract}
KLOE has collected about 30 $pb^{-1}$ in year 2000 at the \dafne collider, which yields the largest
population of $\phi$ meson radiative decays studied so far.
We present the results obtained for \fietag and
\fietapg: the ratio of these two BR's has been measured to be $(5.3\pm 0.5\;
\pm 0.3)\cdot 10^{-3}$, leading to a very accurate determination of the
mixing angle in the flavor basis $\varphi_P=(40^{+1.7}_{-1.5})^{\circ}$ and
to the most accurate determination of BR(\fietapg) to date: $( 6.8 \pm 0.6 \; \pm 0.5 ) \cdot 10^{-5}$. 
\end{abstract}

\section{Introduction}
Radiative decays of light vector mesons to pseudoscalars  have been used as a useful testing ground 
since the early days of the quark model \cite{BeMor65}.
The branching ratio (BR) of the decay \fietapg is particularly interesting 
since its value can probe the $\left|s\bar{s}\right>$ and gluonium content
of the \etap \cite{Close92}. In particular, the ratio of its value to the one
of \fietag can be related to the \Eta-\etap mixing parameters
\cite{Ros83,BraEsSca97,BraEsSca99,BraEsSca01,Feld00} and determine the
mixing angle in the flavor basis \fip , which has been pointed out
as the best suited parameter for a process-independent
description of the mixing. 
In fact, within the two mixing-angles scenario which has emerged from
an Extended Chiral Perturbation Theory framework \cite{KaisLeut98}, as well as from phenomenological
analyses \cite{EsFre99}, has been demonstrated that the two mixing
parameters in the flavor basis are equal apart from terms which violate the
Okubo-Zweig-Iizuka
(OZI) rule \cite{FeldKroll98,Defazio00}, and is thus safe to use one single mixing angle in this basis. 
The
measurements available to date on BR(\fietapg) have still rather large
statistical uncertainties \cite{PDG,ICHEP00}. 
The present analysis of \fietapg decays is  based on an
integrated  luminosity of  $\simeq 17 \pbinv$ corresponding to  about 60\% of
the luminosity collected by the KLOE detector
\cite{kloe} at the \dafne \cite{dafne} $e^+e^-$ collider in Frascati during year 2000.
The accuracy we obtain is significantly better than the current world
average and allows us to extract the \Eta-\etap mixing angle in the flavor
basis with an error of $\simeq 1.5^{\circ}$ from this single measurement.

\section{Analysis}
We use the  following decay chains to
determine the ratio $R= BR(\fietapg)/BR(\fietag)$:
\bi
\item [$\ast$]$\fietapg$;
\bi 
\item[] $\etap\to\pip\pim\eta$;
\bi 
\item[]~~~~~~~~~~~$\eta\to\gamma\gamma$
\ei
\ei

\item [$\ast$]$\fietag$;
\bi 
\item[]$\eta\rightarrow\pip\pim\piz$;
\bi
\item[]~~~~~~~~~~~$\piz\to\gamma\gamma$
\ei
\ei\ei
The final state is $\pip\pim\gamma\gamma\gamma$ for
both the \fietapg and \fietag events, and thus most of the
systematics approximately cancel out when evaluating the ratio $R$; moreover
since the \fietag decays can be quite easily selected with small
background they constitute a very clean control sample for the analysis. 
The \fietag events, being about two orders of magnitude more abundant than the
corresponding \fietapg ones, constitute also the main source of background
for the \fietapg detection.
Further background events can rise from:
\bi
\item[$\ast$] \fikskl events with one charged vertex where
at least one photon is lost and the \Kl is decaying near the interaction
point (IP);
\item[$\ast$] $\phi\rightarrow \pip \pim \piz$ events with an additional
photon detected due to accidental photons or splitting of clusters in the
electromagnetic calorimeter (EmC). 
\ei
\subsection{Event selection: first level}
The events are reconstructed using the standard KLOE reconstruction
libraries and selected from the radiative stream where very loose
cuts are applied to reduce background from \fikskl,  machine background and
cosmic rays events.
Then a first level topological selection for the \pip\pim\phot\phot\phot channel runs as follows:
\begin{itemize}
\item [$\ast$]3 and only 3 prompt neutral clusters (see below) with
$21^{\circ}<\theta_{\gamma}<159^{\circ}$;
\item [$\ast$]Opening angle between each couple of photons $> 18^{\circ}$
\item [$\ast$]1 charged vertex inside the cylindrical region $r<4\;\cm$; $|z|<8 \;\cm$.
\end{itemize}
This selection is common to both \fietapg and \fietag events. 
A prompt neutral
clusters is defined as a cluster in the EmC with no associated track coming
from the Drift Chamber (DC) and 
$|(t-\frac rc)|< 5\sigma_t$ where $t$ is the arrival time on the EmC, $r$ is
the distance of the cluster from the IP and $c$ is the
speed of light and $\sigma_t = 54\ps/\sqrt{E(\GeV)} \oplus 147 \ps
$\cite{nimEmC} where the constant term includes the effect of the source length.

The region below $21^{\circ}$ is excluded due to the presence of \dafne
magnetic quadrupoles near the interaction point. The cut on the opening
angle between photons strongly reduce the effect of cluster splitting.

The overall ``common'' selection efficiencies (including trigger,
reconstruction and first level selection) are 45.8\% and 49.6\%
respectively for \fietapg and \fietag events. The ratio of the two efficiencies
$\varepsilon_{\etap\gamma}/\varepsilon_{\eta\gamma}= 0.923$ 
is reasonably close to one, as expected.
The main reason for this ratio not being one is the difference in
the efficiency to find a charged vertex inside the cylindrical region
around IP; this in turn is related to a slightly different momentum
spectrum of charged pions in the two categories of events.
%the charged pions in \fietapg events have, on average, lower momentum they are more
%likely to spiralize and/or decay inside the DC leading to poorer vertex
%efficiency and resolution with respect to the one in \fietag events.
%In fact, apart from all other effects, Monte Carlo shows that the efficiency
%o find a vertex in the selection region is 62.0\% for \fietag and 57.7\%
%for \fietapg events. The (small) remaining discrepancy is due to the
%slightly different photon energy spectrum for the two categories of events. 

After this selection we perform a kinematic fit constraining
global energy-momentum conservation and the speed of light for each photon,
without imposing any intermediate particle mass constraint.
A loose cut on ${\cal P}(\chi^2)>1\%$ for this fit is imposed for both
\fietag and \fietapg events to ensure the good reconstruction of the event.   
Background from $\phi\to\pip\pim\piz$  events is strongly 
reduced by means of a cut on the charged pions energy endpoints respectively:
\bi
\item[$\ast$] $E_{\pip}+E_{\pim}< 550 \MeV$ (\fietag events)
\item[$\ast$] $E_{\pip}+E_{\pim}< 430 \MeV$ (\fietapg events)
\ei

%Th remaining analysis is mainly devoted to select \fietapg events from
%\fietag ones.
\subsection{\fietapg events selection}
Further selection of \fietapg events is made via a cut on the
total photon energy (to scale down \fikskl background):
\bi
\item[$\ast$]$\Sigma_{\gamma}E_{\gamma} > 540 \MeV$. 
\ei

%\begin{itemize} 
%\item For \fietag selection:
%       \begin{itemize}
%       \item[$\ast$] $E_{\pip}+E_{\pim} < 550 \MeV$ 
%       ($\varepsilon_{3\pi}\simeq 1.5\cdot 10^{-3}$)
%       \item[$\ast$] $E_{\gamma}^{\rm max} > 300 \MeV $     
%       ($\varepsilon_{\Kl\Ks}\simeq 2\cdot 10^{-4}$)
%\end{itemize}
%\item For \fietapg selection:
%       \begin{itemize}
%       \item[$\ast$] $E_{\pip}+E_{\pim} < 412 \MeV$ 
%       ($\varepsilon_{3\pi}\simeq 1 \cdot 10^{-4}$)
%       \item[$\ast$] \Sum_{\gamma}$E_{\gamma} > 520 \MeV$ 
%       ($\varepsilon_{\Kl\Ks}\simeq 1 \cdot 10^{-4}$)
%       \end{itemize}
%\end{itemize}

Contamination from \fietag events into the \fietapg sample is at this level
still very high since about 35\% of \fietag events are still in the sample
(S/B $\approx 5\cdot 10^{-3}$) while contamination from \fikskl and
$\fipippimpiz$ is expected to be small. In fact no Monte Carlo generated event from \fikskl and
$\fipippimpiz$ survives these cuts giving rise to upper limits on these
backgrounds (dominated, in the present analysis, by limited Monte Carlo statistics) given by, respectively:
\bi
\item[$\ast$] $N_{\Ks\Kl} < 6\cdot 10^{-7} \cdot N_{\phi}$ at 90\% C.L. 
\item[$\ast$] $N_{\pip\pim\piz} < 2\cdot 10^{-7} \cdot N_{\phi}$ at 90\% C.L. 
\ei

To select \fietapg events over the \fietag background we exploit the
kinematical properties of the three photon in both categories of events.
The energy spectrum of the photons gives no combinatorial problem : 
radiative photon is the hardest one in \fietag events, while it is
the softest in \fietapg events; the other two photons being generated in
\piz and $\Eta$ decays respectively (fig. \ref{espec}).
If we plot the energy of the two hardest photons (chosen at random to be
``$E_1$'' and ``$E_2$'') after kinematic fit we see (fig. \ref{etapsel})
that a strong correlation between the two photons from \Eta is present in
\fietapg events while the \fietag events are grouped into two bands around
$E_{1(2)}=363$ MeV as expected from the presence of the nearly
monochromatic radiative photon. 
The selection of \fietapg events is then made cutting on an elliptic
shaped region in the $E_1-E_2$ plane whose parameters are:
\bi
\item[$\ast$] Coordinates of the ellipse centre = (285,285) MeV
\item[$\ast$] Major axis = 90 MeV
\item[$\ast$] Minor axis = 25 MeV
\item[$\ast$] Major axis inclination w.r.t. $E_1$ axis = $135^{\circ}$
\ei  
%%%%%%%%%%%%%%%%%%%%%%%%%%%%%%%%%%%%%%%%%%%%%%%%%%%%%%%%%%%%%%%%%%%Figura 1
\begin{figure}[ht]
\bc
\epsfig{file=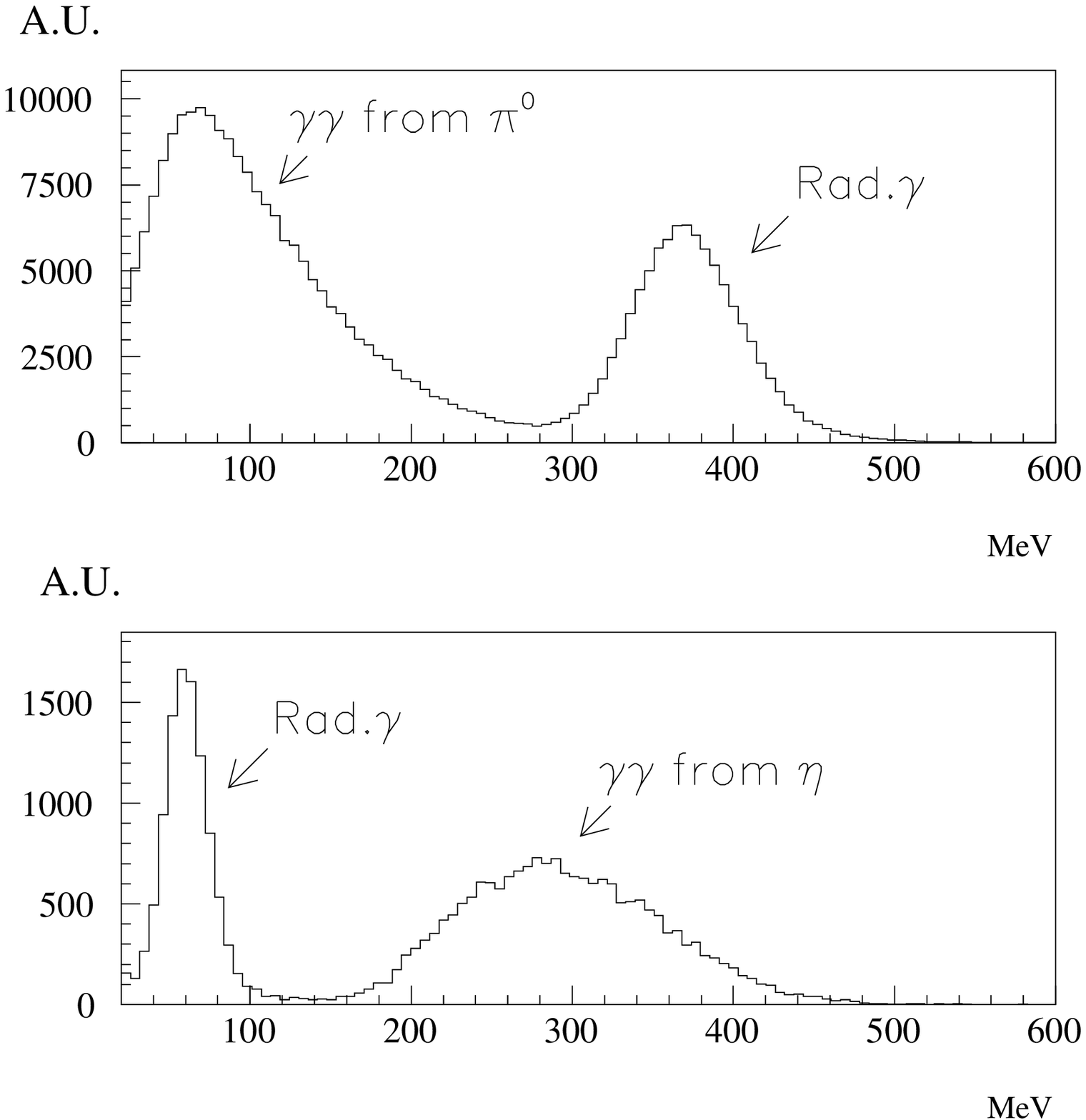,width=8cm}
\caption{{\footnotesize Monte Carlo photon energy spectrum for
\fietag\to\pip\pim\phot\phot\phot (upper plot)  and
\fietapg\to\pip\pim\phot\phot\phot (lower plot).}}
\label{espec}
\ec
\end{figure}
%%%%%%%%%%%%%%%%%%%%%%%%%%%%%%%%%%%%%%%%%%%%%%%%%%%%%%%%%%%%Fig2
\begin{figure}[ht]
\bc
\begin{tabular}{cc}

\epsfig{file=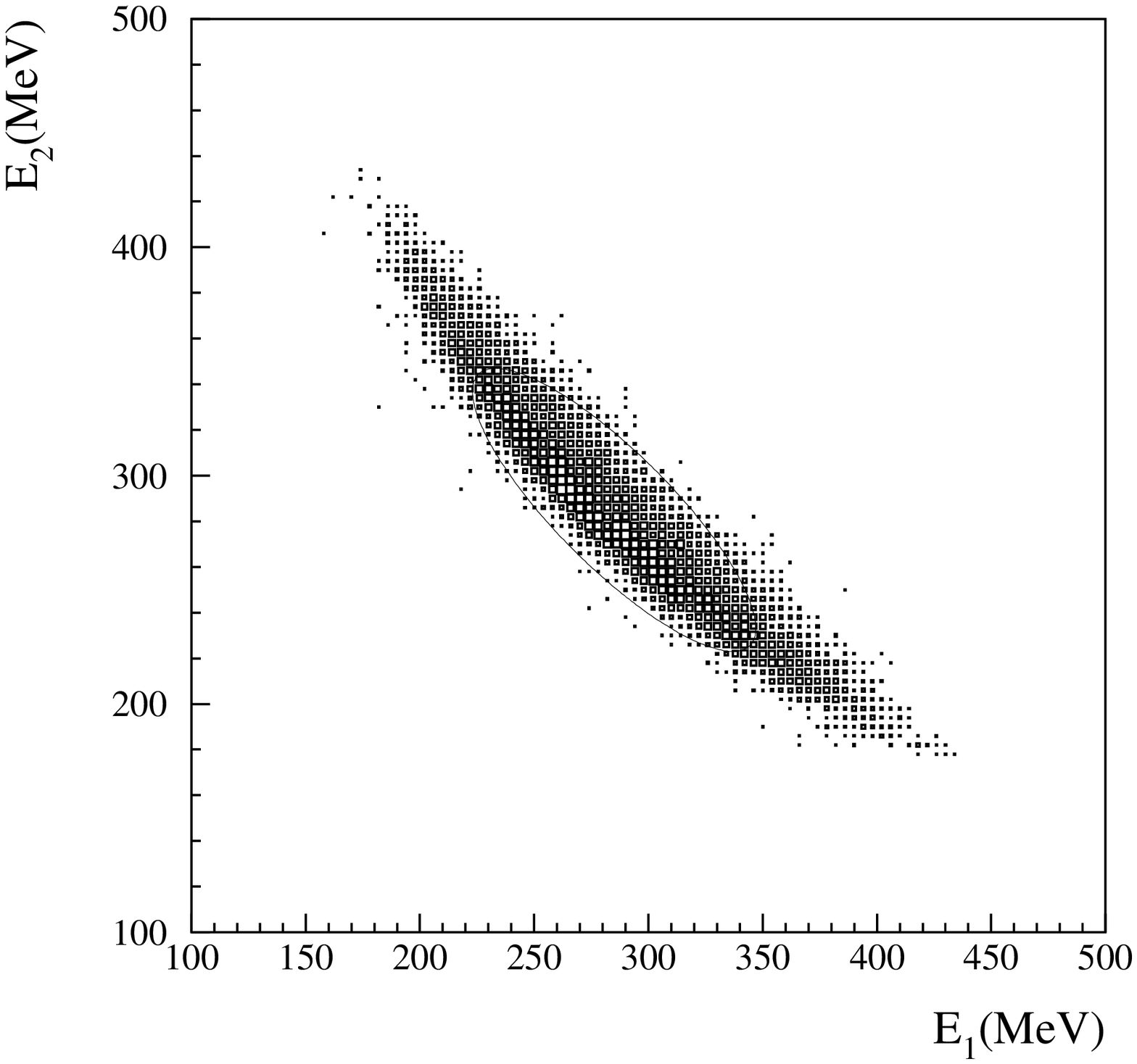,width=8cm}&
\epsfig{file=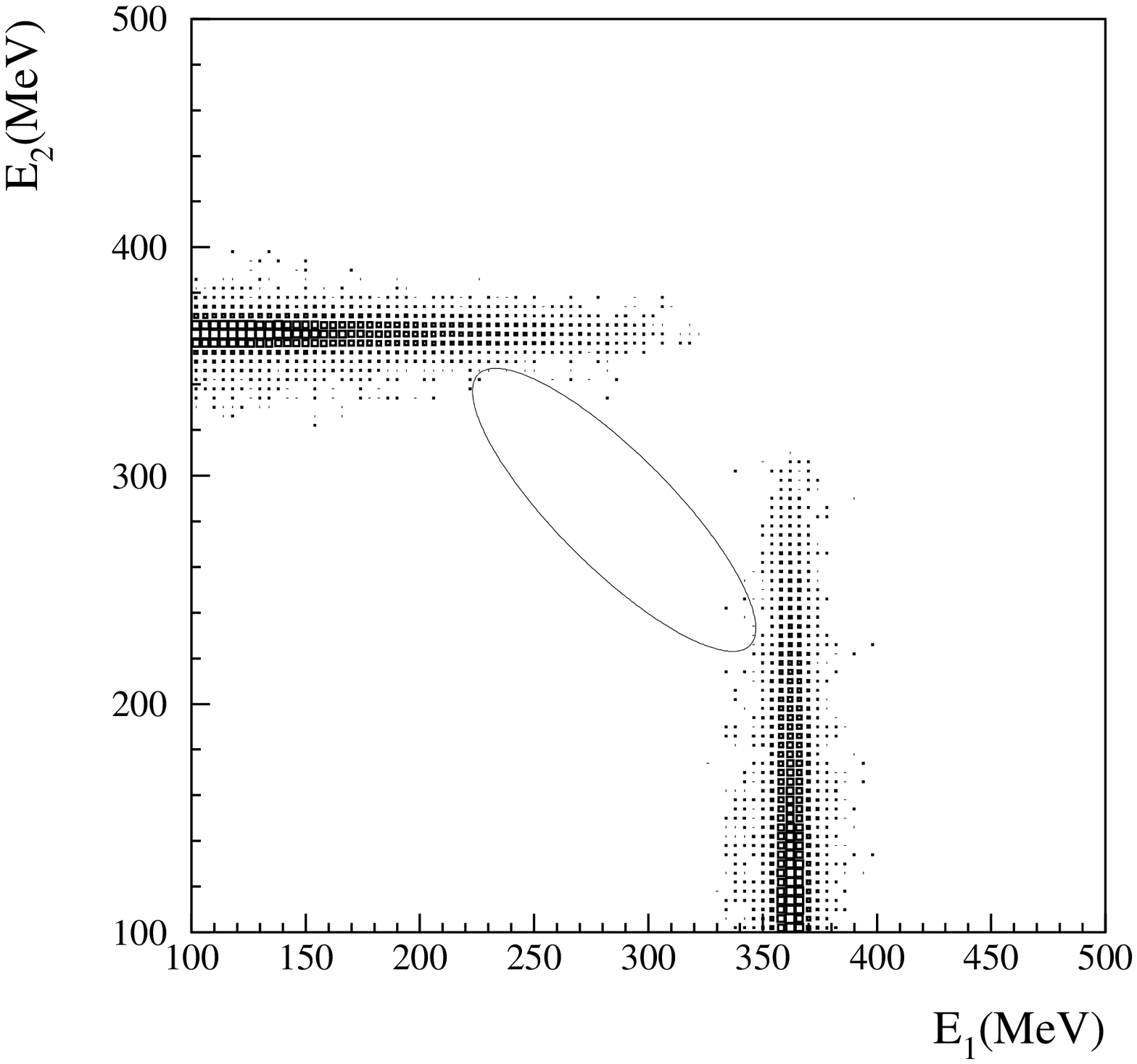,width=8cm}\\
%\multicolumn{2}{c}{
%\epsfig{file=e1vse2last.eps,width=8cm}}\\
\end{tabular}
\caption{{\footnotesize Left: Monte Carlo \fietapg events in the
$E_1-E_2$ plane;
Right: Monte Carlo \fietag events in the $E_1-E_2$ plane. The elliptical
selection region is shown.}}
\label{etapsel}
\ec
\end{figure}
%%%%%%%%%%%%%%%%%%%%%%%%%%%%%%%%%%%%%%%%%%%%%%%%%%%%%%%%%%%%%%%%%%%%%%%%%%%%
If we plot the $\pip\pim\gamma\gamma$ invariant mass for the events inside
the selection ellipse on data we notice a clear peak at the $\etap$
mass value with the same $\sigma$ of the one expected from Monte Carlo,
over a small residual background (see fig.\ref{etap_peak}). To better understand the
shape of background on data a ``donut shaped'' region around the selection
ellipse has been used, and the shape for $M_{\pip\pim\gamma\gamma}$ has
been normalised to the expected Monte Carlo background from \fietag events
(shaded area). Also, a fit to a gaussian plus polynomial
background has been performed to evaluate background directly on data.
Finally the signal has been
selected in the region  $942 \MeV/c^2 \leq M_{\pip\pim\gamma\gamma} \leq 974
\MeV/c^2$ and the expected background subtracted.
The final number of selected events is then $N_{\etap\gamma}=124 \pm 12
({\rm stat.}) \pm 5 ({\rm syst.})$ where the statistical and systematic error includes the
one in estimating background absolute level and shape. Monte Carlo efficiency for this selection on
\fietapg events is $\varepsilon_{\etap\gamma}= 23.0\%$
\clearpage  
%%%%%%%%%%%%%%%%%%%%%%%%%%%%%%%%%%%%%%%%%%%%%%%%%%%%%%%%%%%%%%%%%%%Figura 2
\begin{figure}[ht]
\begin{center}
\epsfig{file=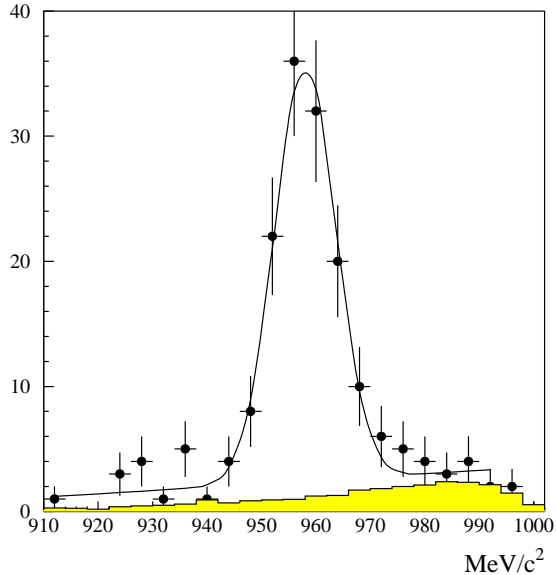,width=9cm}
\end{center}
\caption{{\footnotesize The \pip\pim\phot\phot invariant mass for events selected as
\fietapg candidates. The shaded area is the shape of background obtained selecting
events around the elliptical region and normalised to the expected Monte
Carlo number of events. The continuous line is the result of a gaussian plus
linear fit.}}
\label{etap_peak}
\end{figure}
%%%%%%%%%%%%%%%%%%%%%%%%%%%%%%%%%%%%%%%%%%%%%%%%%%%%%%%%%%%%%%%%%%%%%%%%%%%%
\subsection{\fietag events selection}
Apart from the first level selection, the only additional cut to select
\fietag events is a $\pm 10\;\sigma$ cut on the energy of the radiative photon
(after kinematic fit); we require thus:
\bi
\item[$\ast$] $320 \MeV < E_{\gamma}^{{\rm rad.}} < 400 \MeV$ 
\ei
This cut has almost 100\% efficiency on the signal and is very effective in
reducing residual background from \fikskl events where the endpoint for
photon energies is at 280 \MeV. 
After this cut we are left with $N_{\eta\gamma} = (502.1 \pm 2.2)\cdot
10^2$ events, and the overall efficiency in detecting \fietag events is
evaluated from Monte Carlo to be 37.6\%.
The abundant and pure  \fietag events can be used as control sample to evaluate
systematic effects on the efficiency by comparing data versus Monte Carlo
distributions for the variable on which the cuts are set. 
Some of these
comparisons are shown in fig. \ref{fig_datavsmc}.

%%%%%%%%%%%%%%%%%%%%%%%%%%%%%%%%%%%%%%%%%%%%%%%%%%%%%
\begin{figure}[ht]
\bc
\begin{tabular}{cc}
\epsfig{file=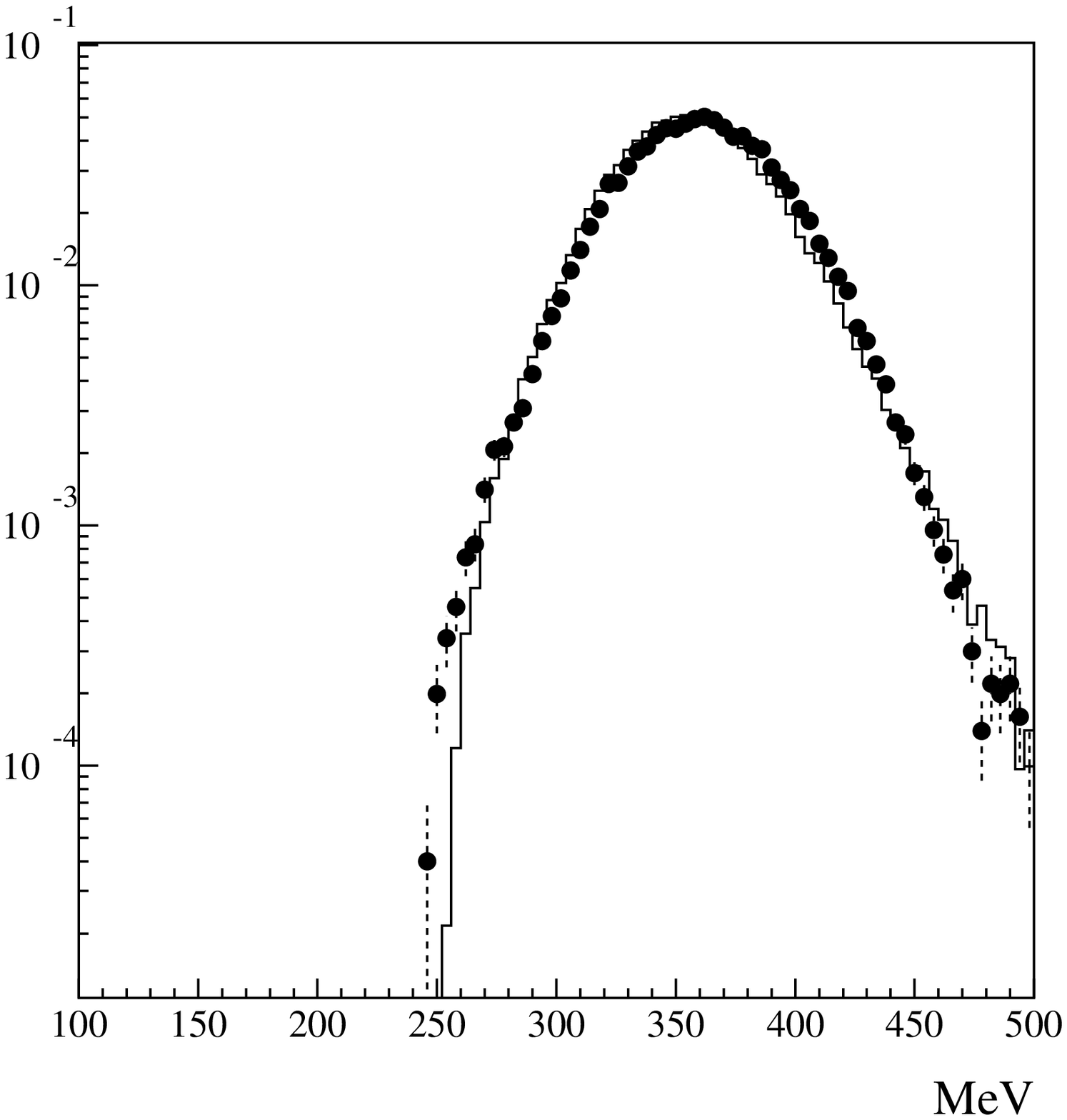,width=65mm}
&
\epsfig{file=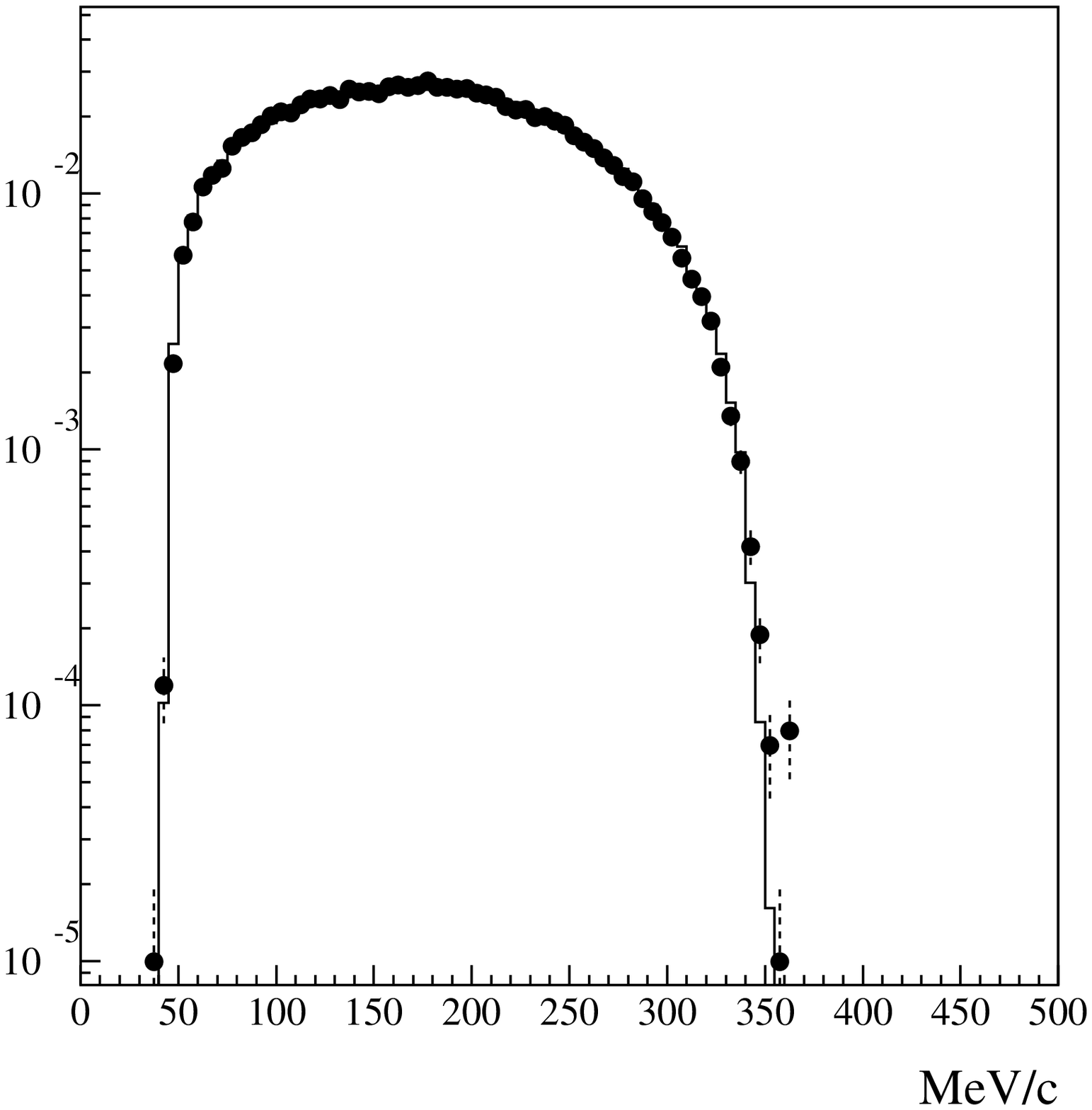,width=65mm}\\
\epsfig{file=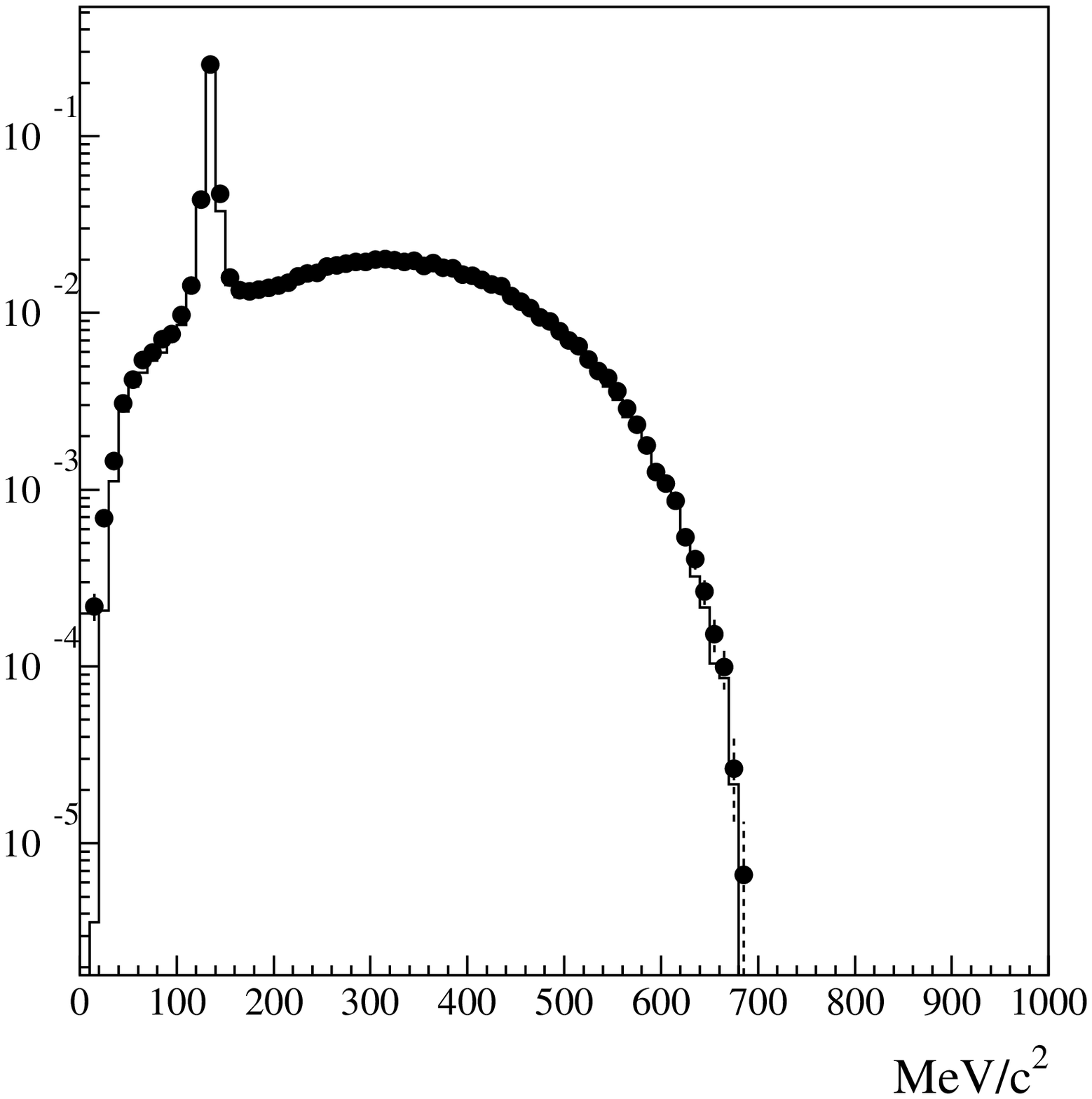,width=65mm}
&
\epsfig{file=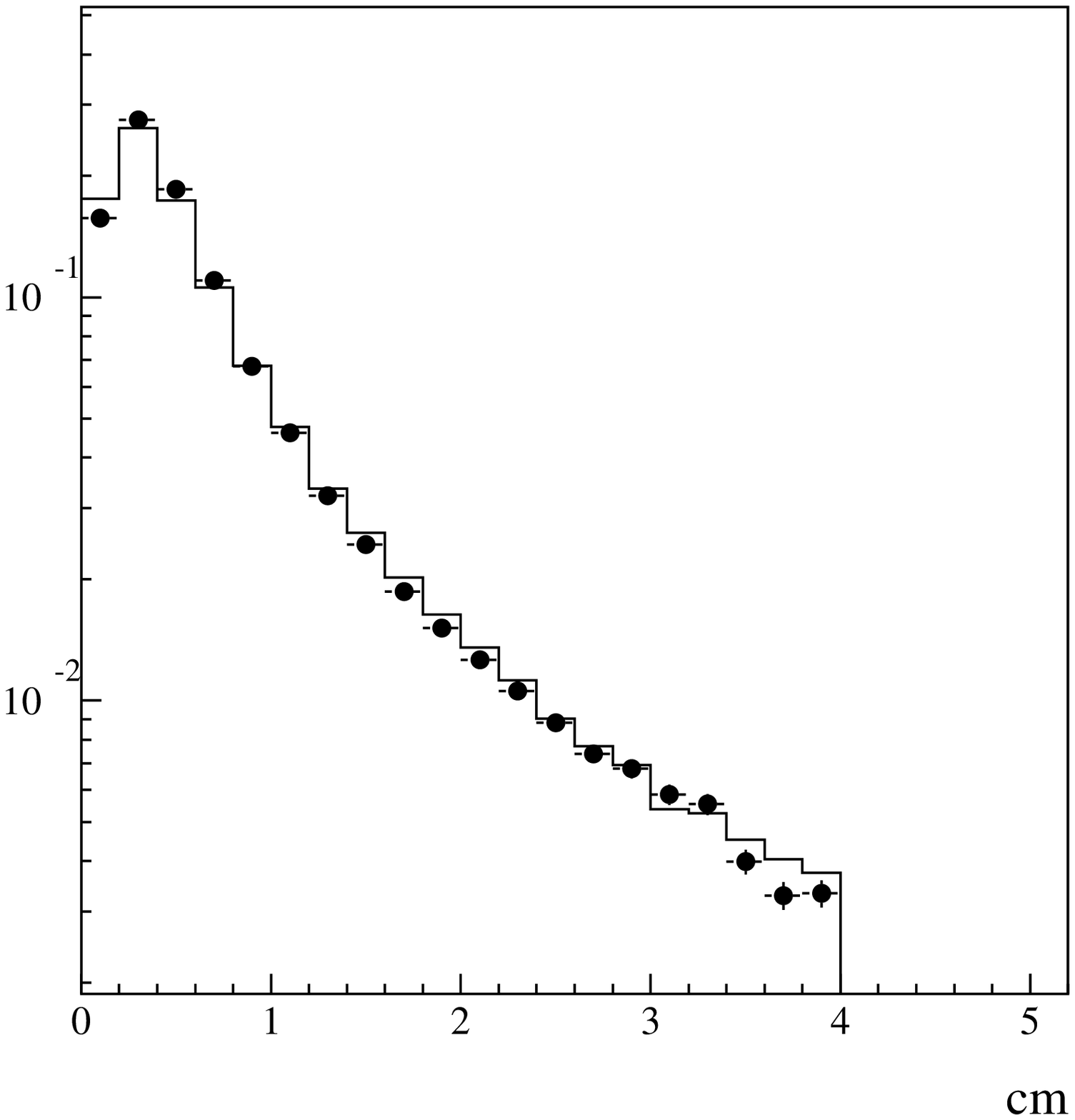,width=65mm}\\
\end{tabular}
\caption{{\footnotesize Monte Carlo pure \fietag (line) versus data (dots) for $\pip\pim\phot\phot\phot$
events selected as \fietag: from top left in clockwise direction
radiative photon energy; charged pions momentum; $\gamma\gamma$ invariant
mass (all combinations); charged vertex position in xy plane. The
histograms are arbitrarily normalised.}}
\label{fig_datavsmc}
\ec
\end{figure}
%%%%%%%%%%%%%%%%%%%%%%%%%%%%%%%%%%%%%%%%%%%%%%%%%%%%%

All comparisons show very good agreement
(at percent level or better) between
data and Monte Carlo:
% and since the dependence of the efficiency on the
%cuts is not critical (for example moving the cut on the charged pions
%energy by $\pm 1\%$ changes \fietag selection efficiency by $\sim 0.1\%$,
%for a more detailed discussion see \cite{tesiPhDFabio})
the overall systematic errors on efficiencies evaluated with Monte Carlo
are thus small.
\clearpage

% Also, when evaluating the ratio $R_{\phi}$ most of the
%systematics will cancel out due to the strong similarities between the two
%categories of events.
%With the statistics of $\sim 2.4 \pbinv$ of 1999 run we found 21 \fietapg
%events in this decay chain with less than one event of background expected at 90\%
%CL, while with the \fietag selection selects 6696 events in the same runs.
%The distribution of the invariant mass of the two charged pions and the two
%most energetic photons in the event is shown in fig. \ref{etapinvmass}
%compared to the Monte Carlo expected for pure \fietapg events.

\section{Results}
The ratio on the number of events selected as \etap\phot and \Eta\phot
respectively, can be related to the ratio of the branching fractions
$R= BR(\fietapg)/BR(\fietag)$ as follows:

$$
R = \frac{N_{\etap\gamma}}{N_{\eta\gamma}}\left(\frac{\varepsilon_{\eta\gamma}}{\varepsilon_{\etap\gamma}}\right)_{common}\times\left(\frac{\varepsilon_{\eta\gamma}}{\varepsilon_{\etap\gamma}}\right)_{analysis}\times\frac{BR(\etapippimpiz)BR(\piz\rightarrow\gamma\gamma)}{BR(\etappippimeta)BR(\etagg)}        
$$
and, thus, using for all quantities in the above formula
the values of table \ref{tabsyst},
where we summarise also the contributions to the final systematic error:
$$
R = \left(5.3 \pm 0.5 (\rm stat.) \pm 0.3 (\rm syst.)\right)\cdot 10^{-3} 
$$

%%%%%%%%%%%%%%%%%%%%%%%%%%%%%%%%%%%%%%%%%%%%%%%Table 1
\begin{table}[h]
\bc
\begin{tabular}{|l|l|l|}
\hline
Quantity & Value & Syst. err. \\\hline
$N_{\etap\gamma}/N_{\eta\gamma}$&$2.5\cdot 10^{-3}$ &4\%\\\hline
$\left(\frac{\varepsilon_{\etap\gamma}}{\varepsilon_{\eta\gamma}}\right)_{common}$&0.923
& $< 1\%$\\\hline
$\left(\frac{\varepsilon_{\etap\gamma}}{\varepsilon_{\eta\gamma}}\right)_{analysis}$&0.662&5\%\\\hline
$\frac{BR(\etapippimpiz)BR(\piz\rightarrow\gamma\gamma)}{BR(\etappippimeta)BR(\etagg)}$&1.30&5\%\\

\hline
\end{tabular}
\caption{{\footnotesize Contributions to the systematic error on $R$. The
5\% systematics on the ratio of analysis efficiencies is evaluated using the \fietag
control sample. The intermediate BR's are taken from \cite{PDG}.}}
\label{tabsyst}
\ec
\end{table}
%%%%%%%%%%%%%%%%%%%%%%%%%%%%%%%%%%%%%%%%%%%%%%%%%%%%%%%%
%\clearpage
This value for $R$ can be related directly to the mixing angle in
the flavor basis. 
In the 
approach by Bramon {\it et al.} \cite{BraEsSca99} where SU(3) breaking is taken into account via
a constituent quark mass ratio $\frac{m_s}{\bar{m}}$ one has:

$$
R = \frac{BR(\phi\rightarrow\eta'\gamma)}
{BR(\phi\rightarrow\eta\gamma)}=\cot^2\varphi_P\left(1-\frac{m_s}{\bar{m}}\frac{\tan\varphi_V}{\sin
    2\varphi_P}\right)^2 \left(\frac{p_{\eta'}}{p_{\eta}}\right)^3
$$

\noi In the approach by Feldmann \cite{Feld00}
chiral anomaly predictions for $P\to\gamma\gamma$ are combined with vector
dominance to extract the couplings $g_{\phi\eta\gamma}$ and
$g_{\phi\etap\gamma}$ which yields, apart from OZI rule violation terms:

$$
R = \frac{BR(\phi\rightarrow\eta'\gamma)}
{BR(\phi\rightarrow\eta\gamma)}=\left(\frac{\sin\fip\sin\varphi_V}{6f_q}-\frac{\cos\fip}{3f_s}\right)^2 \slash
\left(\frac{\cos\fip\sin\varphi_V}{6f_q}+\frac{\sin\fip}{3f_s}\right)^2 \left(\frac{p_{\eta'}}{p_{\eta}}\right)^3
$$

\noi In both cases we use the
result in the cited papers for all parameters entering the ratio except the mixing
angle, in order to estimate the effect of our measurement on the
angle \fip. We get the same
result in extracting the mixing angle in both approaches, i. e. 
$$
\fip = \left(40^{+1.7}_{-1.5}\right)^{\circ}
$$
\noi which would result in a mixing angle in the octet-singlet basis $\tp =
\left(-14.7^{+1.7}_{-1.5}\right)^{\circ}$.
Moreover, using the value in \cite{PDG} for the BR(\fietag) we can
extract the most precise determination of BR(\fietapg) to date (see fig. \ref{bretap}):
$$
BR(\fietapg) =\left( 6.8 \pm 0.6 \;({\rm stat.}) \pm 0.5 \;({\rm
syst.})\right) \cdot 10^{-5} 
$$
\noi This result, given also the value of the mixing angle,  disfavours
large gluonium contents of the \etap \cite{Close92,Ros83,DeshEil80}.
%%%%%%%%%%%%%%%%%%%%%%%%%%%%%%%%%%%%%%%%%%%%%%%%%%%%%
\begin{figure}[ht]
\bc
\epsfig{file=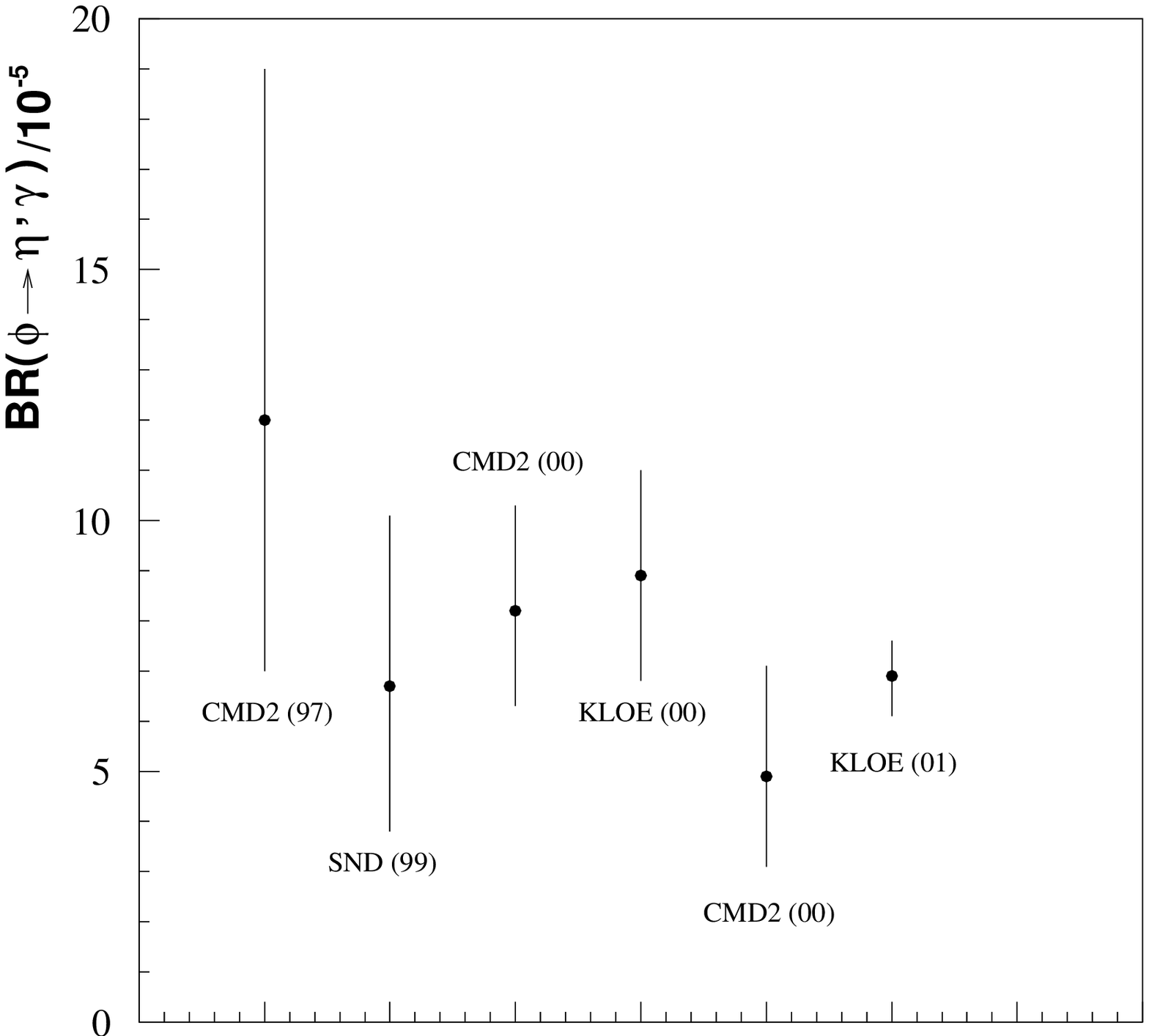,width=80mm}
\caption{{\footnotesize Determinations of the BR(\fietapg) in literature:
CMD2\cite{CMD297,CMD200,CMD200b}; SND\cite{SND99};KLOE(00)\cite{ICHEP00}; KLOE(01) : this work.}}
\label{bretap}
\ec
\end{figure}
%%%%%%%%%%%%%%%%%%%%%%%%%%%%%%%%%%%%%%%%%%%%%%%%%%%%%

\section{Summary}
We present an analysis of about 17 \pbinv of
integrated luminosity at the \dafne collider. We obtain
the best determination of the BR for the relatively rare
process \fietapg. 
The ratio of this BR to the \fietag one helps clarifying
the longstanding \Eta-\etap mixing angle puzzle, and its absolute value disfavours a
large gluonium content for the $\etap$ meson.

\end{document}